\documentclass[journal]{IEEEtran}
\usepackage{eurosym}
\usepackage{color}
\usepackage{refcount}
\usepackage[numbers,sort&compress]{natbib}
\usepackage{setspace}
\usepackage{flushend}
\usepackage[caption = false]{subfig}
\usepackage{float}
\usepackage{fixltx2e}
\usepackage[normalem]{ulem} % either use this (simple) or

\usepackage{graphicx}
\DeclareGraphicsExtensions{.pdf, .pdf.bb, .eps}

\ifCLASSINFOpdf

\else
\fi
\usepackage[cmex10]{amsmath}
\def\mathclap#1{\text{\hbox to 0pt{\hss$\mathsurround=0pt#1$\hss}}}

\usepackage{amssymb}
\usepackage{varioref}
\usepackage{slashbox}

\usepackage{amsmath}% http://ctan.org/pkg/amsmath
\usepackage{algorithm,algcompatible}% http://ctan.org/pkg/{algorithms,algorithmicx}
\algnewcommand\algorithmicreturn{\textbf{return}}
\algnewcommand\RETURN{\algorithmicreturn}
\algnewcommand\algorithmicprocedure{\textbf{procedure}}
\algnewcommand\PROCEDURE{\item[\algorithmicprocedure]}%
\algnewcommand\algorithmicendprocedure{\textbf{end procedure}}
\algnewcommand\ENDPROCEDURE{\item[\algorithmicendprocedure]}%
\algnewcommand{\algvar}[1]{{\text{\ttfamily\detokenize{#1}}}}
\algnewcommand{\algarg}[1]{{\text{\ttfamily\itshape\detokenize{#1}}}}
\algnewcommand{\algproc}[1]{{\text{\ttfamily\detokenize{#1}}}}
\algnewcommand{\algassign}{\leftarrow}

\usepackage{makeidx}
\makeindex

\begin{document}

%
% paper title
% can use linebreaks \\ within to get better formatting as desired
% Do not put math or special symbols in the title.
%\title{On the Security of Networked Control Systems in Smart Vehicles: Study of Covert   Attacks on Adaptive Cruise Control}
\title{On the Security of Networked Control Systems in Smart Vehicle and its Adaptive Cruise Control}
%Cyber Attack Prevention through  Zero False-Positive Rule-Set Generation for Industrial Systems Firewalls \\Prevention of Cyber Attacks on Industrial Systems through  Zero False-Positive   Rule-Set Generation for     Firewalls\\
%
% author names and IEEE memberships
% note positions of commas and nonbreaking spaces ( ~ ) LaTeX will not break
% a structure at a ~ so this keeps an author's name from being broken across
% two lines.
% use \thanks{} to gain access to the first footnote area
% a separate \thanks must be used for each paragraph as LaTeX2e's \thanks
% was not built to handle multiple paragraphs
%
\author{ Faezeh~Farivar,~\IEEEmembership {Member,~IEEE}, Mohammad~Sayad~Haghighi,~\IEEEmembership{Senior Member,~IEEE}\\Alireza~Jolfaei,~\IEEEmembership {Senior Member,~IEEE},  Sheng Wen,~\IEEEmembership {Member,~IEEE},

\thanks{
Copyright (c) 2019 IEEE. Personal use of this material is permitted. However, permission to use this material for any other purposes must be obtained from the IEEE by sending a request to pubs-permissions@ieee.org.}
\thanks{F. Farivar is with the Department of Mechatronics and Computer Engineering, Science and Research Branch, Islamic Azad University, Tehran, Iran, Email: f.farivar@srbiau.ac.ir, farivar@ieee.org.}% <-this % stops a space
\thanks{M. Sayad Haghighi  is with the School
of Electrical and Computer Engineering, College of Engineering, University of Tehran, Iran, Email: sayad@ut.ac.ir, sayad@ieee.org (corresponding author).}
\thanks{A. Jolfaei  is with the Department of Computing, Macquarie University, Sydney, NSW 2109, Australia, Email: alireza.jolfaei@mq.edu.au.}
\thanks{Sheng Wen is with the Department of Computer Science and Software Engineering, Swinburne University of Technology, Australia, Email: swen@swin.edu.au.}
}

% note the % following the last \IEEEmembership and also \thanks - 
% these prevent an unwanted space from occurring between the last author name
% and the end of the author line. i.e., if you had this:
% 
% \author{....lastname \thanks{...} \thanks{...} }
%                     ^------------^------------^----Do not want these spaces!
%
% a space would be appended to the last name and could cause every name on that
% line to be shifted left slightly. This is one of those "LaTeX things". For
% instance, "\textbf{A} \textbf{B}" will typeset as "A B" not "AB". To get
% "AB" then you have to do: "\textbf{A}\textbf{B}"
% \thanks is no different in this regard, so shield the last } of each \thanks
% that ends a line with a % and do not let a space in before the next \thanks.
% Spaces after \IEEEmembership other than the last one are OK (and needed) as
% you are supposed to have spaces between the names. For what it is worth,
% this is a minor point as most people would not even notice if the said evil
% space somehow managed to creep in.

% The paper headers
\markboth{}%
{Shell \MakeLowercase{\textit{et al.}}: Bare Demo of IEEEtran.cls for Journals}
% The only time the second header will appear is for the odd numbered pages
% after the title page when using the twoside option.
% 
% *** Note that you probably will NOT want to include the author's ***
% *** name in the headers of peer review papers.                   ***
% You can use \ifCLASSOPTIONpeerreview for conditional compilation here if
% you desire.

% If you want to put a publisher's ID mark on the page you can do it like
% this:
%\IEEEpubid{0000--0000/00\$00.00~\copyright~2012 IEEE}
% Remember, if you use this you must call \IEEEpubidadjcol in the second
% column for its text to clear the IEEEpubid mark.

% use for special paper notices
%\IEEEspecialpapernotice{(Invited Paper)}

% make the title area
\maketitle

% As a general rule, do not put math, special symbols or citations
% in the abstract or keywords.
\begin{abstract}
With the benefits of Internet of Vehicles (IoV) paradigm, come along unprecedented security challenges. Among many applications of inter-connected systems, vehicular networks and smart cars are examples that are already rolled out. Smart vehicles not only have networks connecting their internal components e.g. via Controller Area Network (CAN) bus, but also are connected to the outside world through road side units and other vehicles. In some cases, the internal and external network packets pass through the same hardware and are merely isolated by software defined rules. Any misconfiguration opens a window for the hackers to intrude into  vehicles' internal components e.g.  central lock system,   Engine Control Unit (ECU),  Anti-lock Braking System (ABS) or  Adaptive Cruise  Control (ACC) system. Compromise of any of these can lead to disastrous outcomes. In this paper, we study the security of smart vehicles' adaptive cruise control systems in the presence of covert  attacks. We define two covert/stealth attacks in the context of cruise control and propose a novel intrusion detection and compensation method to disclose and respond to such attacks. More precisely, we focus on the covert cyber attacks that compromise the integrity of cruise controller and employ a neural network identifier in the IDS engine to estimate the system output dynamically and compare it against the ACC output. If any anomaly is detected, an embedded substitute controller kicks in and takes over the control. 
We conducted extensive experiments in MATLAB to evaluate the effectiveness of the proposed scheme in a simulated environment.  
\end{abstract}

% Note that keywords are not normally used for peerreview papers.

% For peer review papers, you can put extra information on the cover
% page as needed:
% \ifCLASSOPTIONpeerreview
% \begin{center} \bfseries EDICS Category: 3-BBND \end{center}
% \fi
%
% For peerreview papers, this IEEEtran command inserts a page break and
% creates the second title. It will be ignored for other modes.
\IEEEpeerreviewmaketitle

\section{Introduction}
% The very first letter is a 2 line initial drop letter followed
% by the rest of the first word in caps.
% 
% form to use if the first word consists of a single letter:
% \IEEEPARstart{A}{demo} file is ....
% 
% form to use if you need the single drop letter followed by
% normal text (unknown if ever used by IEEE):
% \IEEEPARstart{A}{}demo file is ....
% 
% Some journals put the first two words in caps:
% \IEEEPARstart{T}{his demo} file is ....
% 
% Here we have the typical use of a "T" for an initial drop letter
% and "HIS" in caps to complete the first word.

% The very first letter is a 2 line initial drop letter followed
% by the rest of the first word in caps.
% 
% form to use if the first word consists of a single letter:
% \IEEEPARstart{A}{demo} file is ....
% 
% form to use if you need the single drop letter followed by
% normal text (unknown if ever used by IEEE):
% \IEEEPARstart{A}{}demo file is ....
% 
% Some journals put the first two words in caps:
% \IEEEPARstart{T}{his demo} file is ....
% 
% Here we have the typical use of a "T" for an initial drop letter
% and "HIS" in caps to complete the first word.

\IEEEPARstart{I}{ntrusion} Detection Systems (IDS), in their classic definition in the cyber world,   are security systems that monitor  the traffic coming in or going out of a network  \cite{farivardetection}. However, as the systems get more complex and interconnected, intrusions find more sophisticated instances and get out of the cyber-only world. Cyber Physical Systems (CPS) are where the cyber and physical worlds meet. They are usually mixtures of IT systems or networks and  electro-mechanical  entities. Examples of such systems are self-driving cars \cite{alheeti2015intrusion},
remote surgical robots \cite{mitchell2013effect, mitchell2014behavior} and smart power grids \cite{sridhar2011cyber, mohammadali2016novel}. 

Attacks launched on CPSs can cause damages in the real world and that will make their  security a big concern in the coming years. With the realization of Industry 4.0, this threat becomes even bigger. A simple cyber attack on one CPS can create a cascade of failures in a complex interconnected system.  

The previously-existing channels of intrusion, including  computer networks \cite{haghighi2010neighbor}, can open the vulnerability window to the CPS of network-controlled type. There are many instances of such systems around. For example, new vehicles, including smart vehicles, use Controller Area Network (CAN) bus to connect different micro-controllers governing the physical components of a vehicle. With the emergence of Internet of Vehicles (IoV) and Vehicular Ad hoc Networks (VANET) \cite{toorchi2013markov} and connection of smart vehicles to the outside world for e.g. Over The Air (OTA) software updates or information gathering, this  threat has become even more serious \cite{harris2015researcher}. In July 2015, Fiat Chrysler recalled 1.4 million cars  after two  hackers demonstrated how to penetrate and control a Jeep Cherokee remotely by using e.g. the car's entertainment system which is  supposedly connected to the mobile data network~\cite{bbc_hack2015, bbc_hack2015_2}. 
Such incidents drew the attention of researchers to  the security of vehicles' CPSs~\cite{kang2016intrusion}.

The issue is that, more and more vehicles are relying on networked control systems whose collision domains or buses are  shared with (or have a route  to) the units connected to the Internet. The isolation is at best, enforced by some sort of software defined network (SDN) controller. This makes the attack surface larger and any misconfiguration can open a window to the internals of smart vehicle.

 Classic Intrusion Detection Systems (IDS) were built to spot anomalies in the world of computer networks. However, in a CPS,  analyzing the traffic alone without knowing the effect of each packet effect on the physical system, one cannot judge about its maliciousness. 

Reference \cite{de2017covert} has categorized  CPS threats and defined a new family which is so called the “covert” attacks. Covert attacks are a subset of deception attacks in which an attacker takes control of some parts of the CPS, e.g. the control signal over the network channel, but tries to keep the attack effect small enough so that it is not detected by human observers. There are two types of covert attacks; one tries to create  steady-state errors on the output of  physical entity to e.g. exhaust or degrade the service performance over the time and the second type tries to create transient 
effects on the CPS output, perhaps repeatedly over a long period to achieve the same goal.

\begin{figure*}[!htb] %[b] %[p] puts at the end
%\begin{left}
   \centering
    \subfloat[]{\includegraphics[width=1\columnwidth, height=.31\columnwidth]{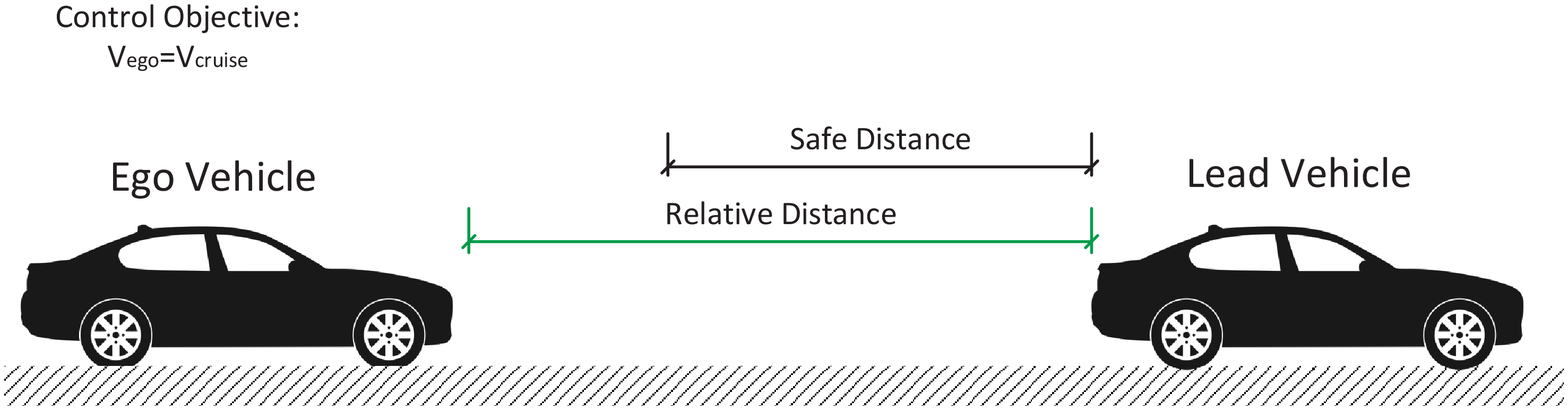}}      \vspace{-2ex}
    \\
     \subfloat[]{\includegraphics[width=1\columnwidth, height=.31\columnwidth]{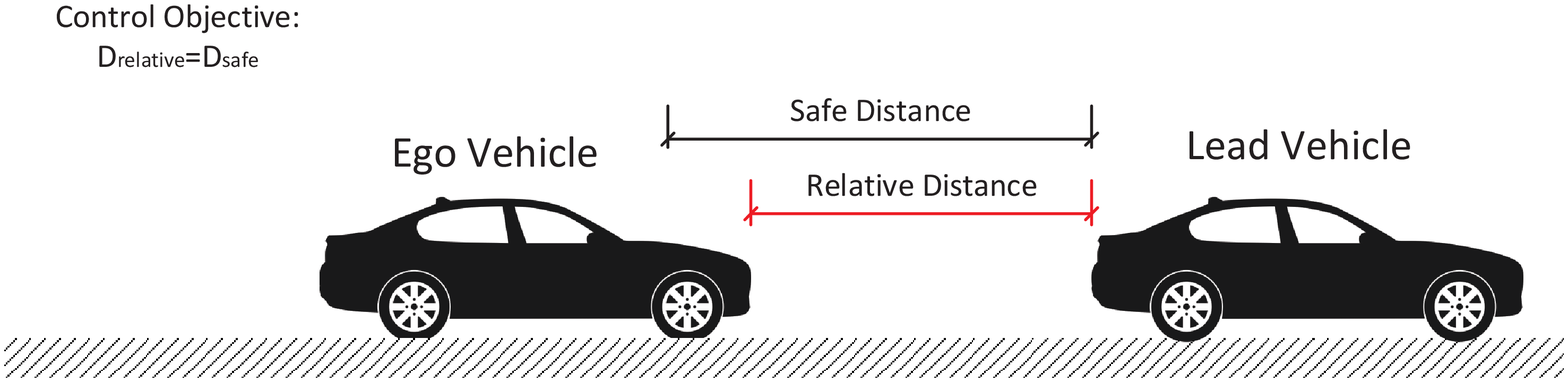}}%
       \vspace{-2ex}
  \\
    \subfloat[]{\includegraphics[width=1\columnwidth, height=.31\columnwidth]{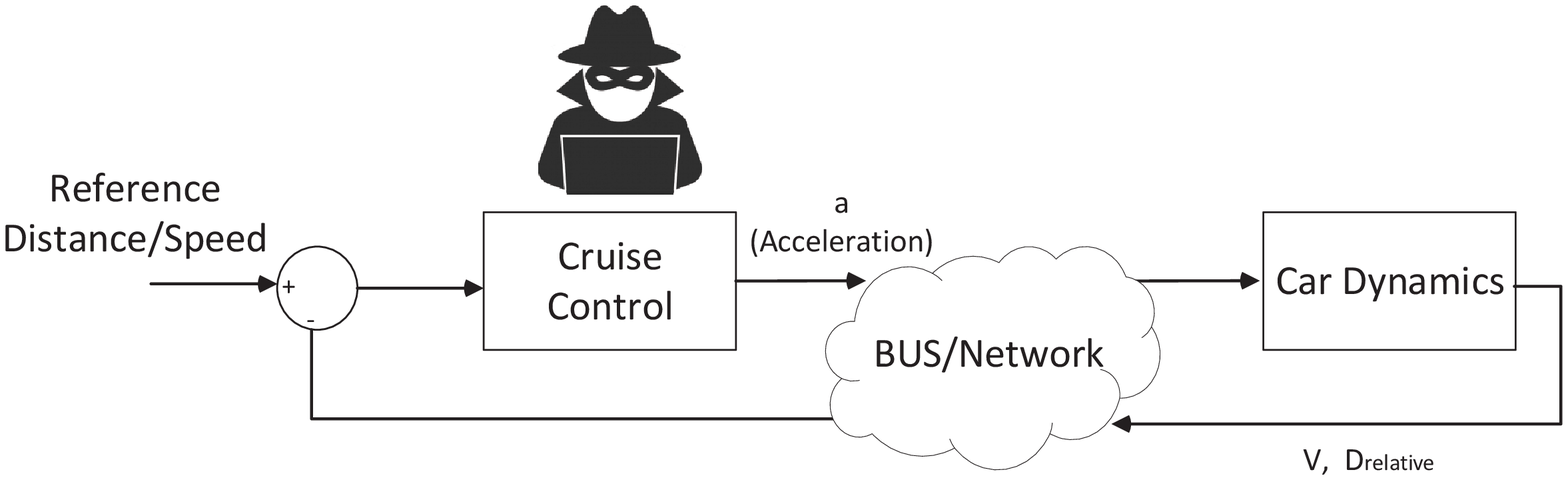}}%

\caption{Demonstration of Adaptive Cruise Control (ACC) in different modes along with the control goals in each case: a) The relative distance is greater than the safe distance, b) the relative distance is smaller than the safe distance. c) The  ACC loop is closed via in-car network that delivers  the control and feedback signals. }
\label{fig:concept}
%\end{left}
\end{figure*}

Inspired by the definition of covert  attacks, this paper proposes two covert attacks on Adaptive Cruise  Control (ACC) systems.
In the first one, the attacker waits for the reference e vehicle to lower its speed and at the right moment and by creating a spike in the acceleration or manipulation of the controller reference, tries to cause an accident or increase its chance. In the second one, the attacker takes over the cruise controller and manipulates the reference signal to lower the distance to the reference car so that the risk of accidental crash is increased. 

We then propose an intrusion detection mechanism for such covert attacks. This IDS takes advantage of a reference model trained by an observer to mimic the system response as closely as possible. A significant deviation from the expected behavior will make the IDS trigger an alarm and switch to the fail-safe controller embedded in the IDS or compensation system. 

In order to evaluate the efficacy of the proposed detection and compensation scheme, we conducted extensive simulations by using Matlab Simulink platform. The results show that the proposed detection method successfully discloses covert attacks on AAC systems and the compensator, which is a simple embedded P controller, also manages to substantially reduce the attack impact on the system compared to when no protection mechanism is employed. 

The rest of this paper is organized as follows: In Section~\ref{section:ACC} the model of the ACC system is described and its control goals are presented.  Section~\ref{section:proposed_model} presents  novel covert attack strategies on the ACC system and their corresponding compensation plans.  Moreover, the IDS system is designed in this section. Section~\ref{section:results} tests the IDS as well as its compensator in a simulation setup and discusses the results.  Related studies are reviewed in Section ~\ref{section:related work}. The paper is concluded in Section~\ref{section:conclusion}.

\section{The Adaptive Cruise Control \label{section:ACC}}
In this section, we  look into the description of the ACC system and its control goals.
An
ACC model is depicted in Fig.~\ref{fig:concept} (a) and Fig.~\ref{fig:concept} (b).
A vehicle, called ego, is assumed to be equipped with ACC system. The ego vehicle has a radar to measure the distance to the front vehicle, which is so called the  lead vehicle. The lead vehicle is presumably  driving in the same lane as the ego vehicle. In addition, the relative velocity is measured by the sensor/radar. 
The control objectives of the ACC system is to make the ego vehicle travel at a driver-specified speed as long as it maintains a safe distance with the lead vehicle. 

There are two modes that the ACC system controls the ego vehicle with. In the first one,  the speed (of the ego vehicle) is set to a driver-set speed as long as a minimum distance is maintained with the lead vehicle. In the second mode, the space between the two vehicles is controlled (by changing the ego vehicle's speed) so that the two vehicles do not get closer than a safe distance. According to real-time  measurements, either  working mode can be enabled by the ACC system. For example, if the relative distance is considerably decreased, the ACC systems switches from mode 1 to mode 2. Likewise, if the distance goes beyond a threshold, the ACC system switches from mode 2 to mode1. To determine the ACC system operating mode, the following rules are considered:

\begin {itemize}
\item
If $D_{rel}\geqslant D_{Safe}$, then the speed control mode is active. Track the driver-set velocity. $V_{set}$ is the goal.

\item
 If $D_{rel}<D_{Safe}$, then the space control mode is active. Keep a  safe distance with the lead vehicle. $D_{Safe}$ is the goal. 
\end {itemize}

\subsection{ACC Mathematical Model}
The model dynamics of the lead and ego vehicles are the same. Let $x$, $v=\dot x$, $a=\dot v=\ddot x$ denote the position, velocity, and acceleration of vehicles, respectively. A simple first order equation expressing  the car cruise control operation is shown below:
\begin{align}
\dot x_{ego}(t)&=F_v(D_{rel}(t))
\\D_{rel}(t)&=x_{lead}(t)-x_{ego}(t)
\label{CarModel_1}
\end{align}
where $F_v$ is a velocity function that explains how the ego vehicle chooses to go, given that its distance to the lead vehicle is $D_{rel}$. The function $F_v(D_{rel})$ can be learned from the sensor data taken along the road, or can be inferred from human driving behaviors. A key modeling weakness of the first order model is that it is not exactly possible for drivers to choose their velocity, because vehicles have inertia due to their mass. Therefore, different velocities can usually be achieved only via gradual acceleration or deceleration. The second order model describes this via Newton's laws of motion:
\begin{align}
\ddot x_{ego}=F_a(x_{lead}-x_{ego},\dot x_{lead}-\dot x_{ego},\dot x_{ego})
\label{CarModel_2}
\end{align}

Here $F_a (D_{rel}, v_{rel}, v_{ego})$ models the acceleration of the ego vehicle as some nonlinear function of the distance to the lead vehicle, the relative velocity to the lead vehicle, and the the ego vehicle velocity. Two second order models are the following-the-leader and the optimal velocity  models. The following-the-leader  model is characterized by: 
\begin{align}
\ddot x_{ego}=\alpha \frac{\dot x_{lead}-\dot x_{ego}}{x_{lead}-x_{ego}}
\label{CarModel_3}
\end{align} 
where $\alpha$ is a constant parameter. The optimal velocity   model is as follows:
\begin{align}
\ddot x_{ego}=\beta(F_v ({\dot x_{lead}}-{\dot x_{ego}})-{\dot x_{ego}})
\label{CarModel_4}
\end{align} 
in which $\beta$ is a constant parameter. In the the following-the-leader model, each vehicle has its own velocity related to its lead vehicle's velocity.
In the optimal velocity model, there is a velocity function $F_v(D_{rel})$ and the ego vehicle velocity relates to that optimal velocity that corresponds to the given spacing $x_{lead}-x_{ego}$. The two models can also be combined as follows:
\begin{align}
\ddot x_{ego}=\alpha \frac{\dot x_{lead}-\dot x_{ego}}{x_{lead}-x_{ego}}+\beta(F_v ({\dot x_{lead}}-{\dot x_{ego}})-{\dot x_{ego}})
\label{CarModel_5}
\end{align} 

As mentioned before, the model dynamics of the lead and ego vehicles are the same. For both ego and lead vehicles, the dynamics between acceleration and velocity in Laplace domain are modeled as:
\begin{align}
G(s)=\frac{1}{s(0.5s+1)}
\label{tf}
\end{align}
which approximates the dynamics of the throttle body and vehicle inertia.
The inputs of the ACC system are the driver-set velocity $V_{set}$,  time gap  $T_{gap}$ which is the time gap between the vehicles,  velocity of ego vehicle $V_{ego}$, relative distance to the lead vehicle $D_{rel}$, and the relative velocity to the lead vehicle $V_{rel}$ (which is  given by the vehicle radar). The output of the ACC system is the acceleration of the ego vehicle.

The safe distance between the lead and ego vehicles is calculated as follows \cite{shakouri2011adaptive}:
\begin {align}
D_{safe}=D_{default}+T_{gap}V_{ego}
\label{Dsafe}
\end {align}
where $D_{default}$ is the standstill default spacing and $T_{gap}$ is the time gap between the vehicles.

\section{Intrusion Detection and Compensation}\label{section:proposed_model}
In this section,  we  assume that the ACC system is prone to covert or stealth attacks which disrupt the control operation. First, we
introduce two attack scenarios for 
 the ACC system. Then,  
we design an intelligent intrusion detection system for  ACC. Then, a novel idea  is presented to compensate the effect of the detected intrusions in both scenarios.
%a false zero classifier as a firewall rule generator. The proposed technique is a batch mode learning for binary classification problems. In the first stage of this study, we focus on the learning technique where there is not  any feature selection on the network intrusion data set. First, main concepts of SVM classifier is briefly described. Then, we explain the proposed technique. In this study, we also take into consideration the selected features which have been determined by a real implementation study of network intrusion data and windows firewall rules.     

\subsection{Covert Attack Scenarios }\label{section:attack scenarios}

% \begin{figure*}[h] %[b] %[p] puts at the end
% %\begin{left}
% %   \centering
%    \subfloat {\includegraphics[width=1\columnwidth, height=0.9\columnwidth]{covert1_IDS_withoutComp.pdf}}
%    \subfloat{\includegraphics[width=1\columnwidth, height=0.9\columnwidth]{covert2_IDS_withComp.pdf}}
%     %   \vspace{-2ex}
% \caption{The intelligent intrusion detection system for the first scenario which the attack time is occurred at  $40$ sec and the detection is the same as the attack time.  (Left) There is not any compensation strategy. (Right) The P Controller is applied to compensate the malicious attack occurred into the ACC system. }
% \label{fig.result_scen2}
% %\end{left}
% \end{figure*}
% 
% 
% 

As mentioned in the previous section, the ACC system operates in two modes; the speed control mode and the space control mode. In this paper, two attack scenarios are studied which target these control objectives of the ACC system. Both attack scenarios are covert by definition. They intrude the ACC system and change  its output. In other words, the attacker disrupts the acceleration of the ego vehicle to change the normal operation of the ego vehicle.
\\

\textbf{Attack Scenario \#1:} The first stealth or covert attack starts with compromising the ACC unit. The attacker remains dormant and monitors the measured distance to the front vehicle while letting the ACC do its job. At the times this distance is at its lowest (presumably near the minimum safe distance), it creates a spike in the control signal and makes the vehicle  accelerate, hoping that by temporally lowering the distance to  something less than the standard or  increasing the speed to something above the limit, the chance of having accidents is increased. This could similarly happen when the front vehicle suddenly brakes and the compromised ACC refuses to reduce the speed. 
\\

\textbf{ Attack Scenario \#2:}  The attacker compromises the ACC unit similar to the first attack, however, unlike the first scenario he/she does not ambush for the attack. The attacker  trivially lowers the ACC's reference distance i.e. $D_{Safe}$ in this scenario. Therefore, during the times ACC is in mode 2 and tries to maintain the safe distance, it is practically following a false reference which gets the ego car closer to the lead vehicle than it really should be. However, since this difference is trivial and not noticeable to the driver, this attack remains covert or stealth. The result is not trivial though. Statistically speaking, depending on the road condition (e.g. wet or dry), the braking power of the lead and ego vehicles, and the ego vehicle's driver reaction time, the chance of having accidents considerably increases.
\\

\begin{figure*}[!htb] %[b] %[p] puts at the end
%\begin{left}
   \centering
   \includegraphics[width=1.8\columnwidth, height=0.8\columnwidth ]{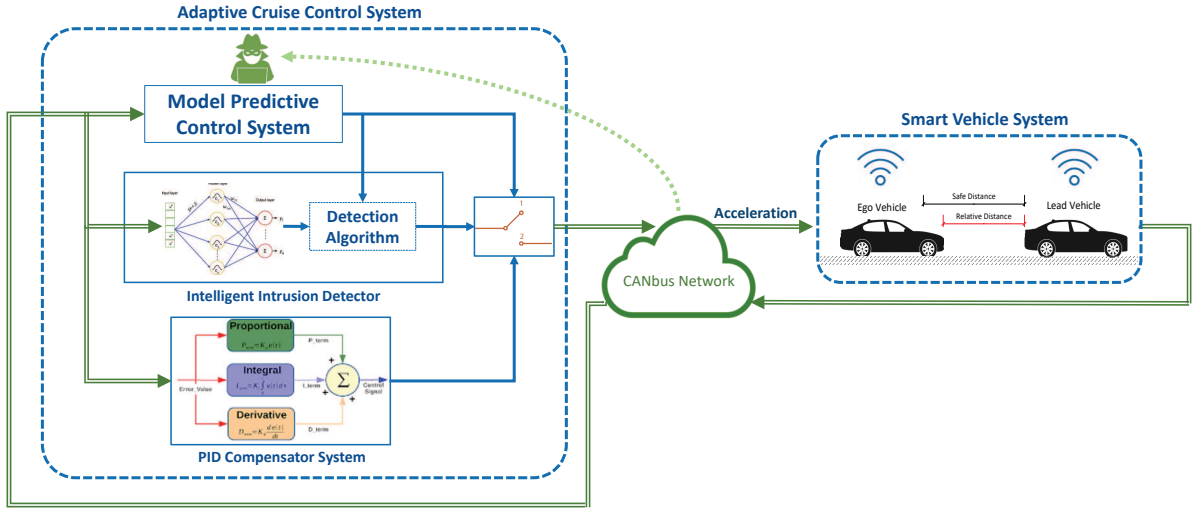}      \vspace{0ex}
\caption{The proposed scheme for detection and compensation of malicious attacks into the adaptive cruise control system. }
\label{fig:scheme}
%\end{left}
\end{figure*}

\subsection{Intelligent Intrusion Detection}
{Accuracy in detection and reaction times are important issues in CPS security. Reaction time is critical to prevent the process failures. Early detections increase the chance for  compensator to maintain CPS performance.}

We propose to use an Intrusion Detection System (IDS) to spot the covert attacks introduced. We assume that there is an initial safe interval during which an identifier learns the CPS/plant dynamics. This could be as simple as a transfer function identification for linear systems. The initial safe period assumption is not unreasonable since a vehicle is not hacked or infected by a worm as soon as it comes out of the factory.  \\  Therefore, we assume that attacks, if there are any, are launched after $T_ {attack}$ seconds. 
An intelligent identifier is used to construct a model of the CPS while it operates in the safe interval. The identifier we employ is an artificial neural network. \\ In this study, we suggest using   adaptive hard thresholds to detect anomalies. The sum of modeling errors and quantization errors (as a result of digitizing the signals and passing them through the network) form a residual error $(Y_{out}-Y_{nn})$ based on which IDS determines the detection threshold. This error  statistical data is collected from the plant  during the safe operation interval. Hypothetically speaking, if the aggregated errors have a Gaussian distribution in the  state $\psi_i$, the alert threshold should be  set to $\mu_i\pm k\sigma_i$~\cite{farivardetection}. Here, $k$ is a constant coefficient and $\mu_i$ and $\sigma_i$ are the  mean and std of the Gaussian error distribution in system state $\psi_i$. According to \cite{farivardetection},   the probability of getting a false positive equals $2Q(k)=erfc(\frac{k}{\sqrt{2}})$. \ Decreasing false positives usually results in an increase of false negatives. Therefore, $k$ should be selected in a way   that  significant deviations are captured and small perturbations are ignored.  The proposed scheme for detection and compensation of the malicious attack in the ACC system is demonstrated in Fig.~\ref{fig:scheme}.

\subsection{Control Systems and Compensation Strategies}

   In this study, we take a model predictive control (MPC) system as the main ACC core.
The  MPC
system controls the  vehicle  in the normal mode. If the IDS that monitors ACC behavior, detects an anomaly, it means that an  attack has happened. Therefore, IDS  switches from  the main control system to the compensator in order to alleviate the effect of  attack (on the  the ego car).   

\begin{figure}[h] %[b] %[p] puts at the end
%\begin{left}
   \centering
   {\includegraphics[width=.9\columnwidth, height=0.34\columnwidth]{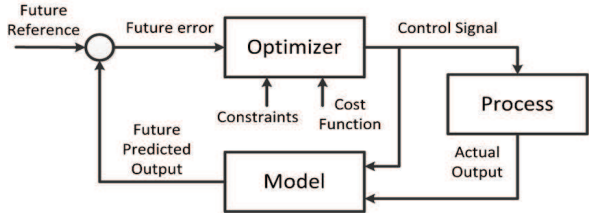}} \vspace{-2ex}
    \\
   % \subfloat[]{\includegraphics[width=0.85\columnwidth,]{PID2.eps}}%
       \vspace{2ex}
 \caption{Model predictive control system \cite{tzovla2000simplified}.}
\label{fig:controller}
%\end{left}
\end{figure}

As mentioned, the core of ACC is an MPC  which  applies the linear model of the system, disturbance and noise models to estimate the state of the control system and  also anticipate the system future outputs. By using the anticipated outputs, the MPC solves a quadratic programming optimization problem to provide optimal  adjustments for variables. The structure of an MPC system is shown in Fig.~\ref{fig:controller}.

The compensator  proposed in this study is a PID control system. The reason for this choice is the popularity of this family of controllers in industry and the low cost of embedding such a unit inside the intrusion detection and compensation system.  In a case study, we design  P controllers to compensate the effect of   covert attacks in the scenarios   introduced before.

\begin{figure*}[!htp] %[b] %[p] puts at the end
%\begin{left}
%   \centering
   \subfloat[] {\includegraphics[width=1\columnwidth, height=1\columnwidth]{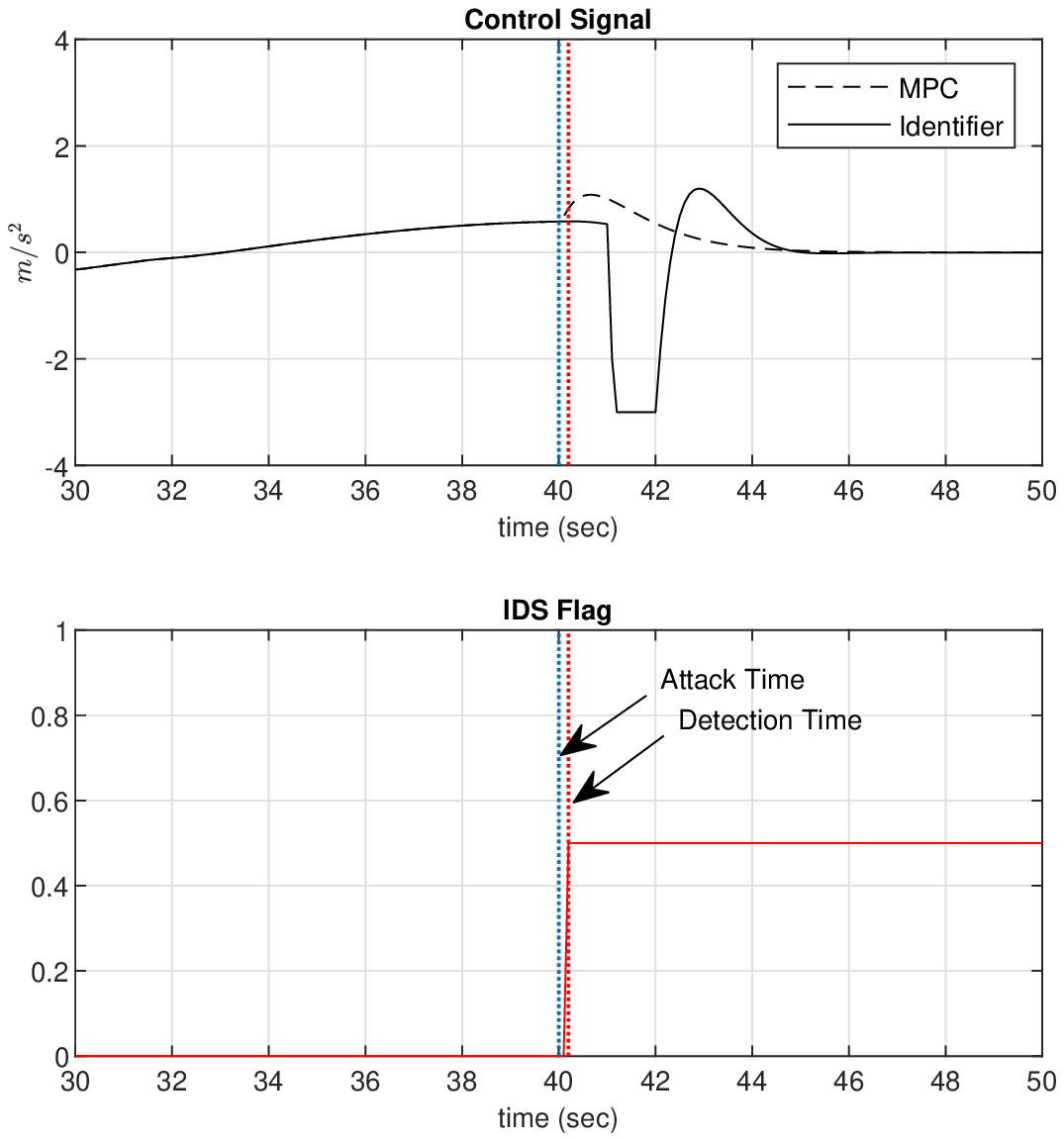}}
   \subfloat[] {\includegraphics[width=1\columnwidth, height=1\columnwidth]{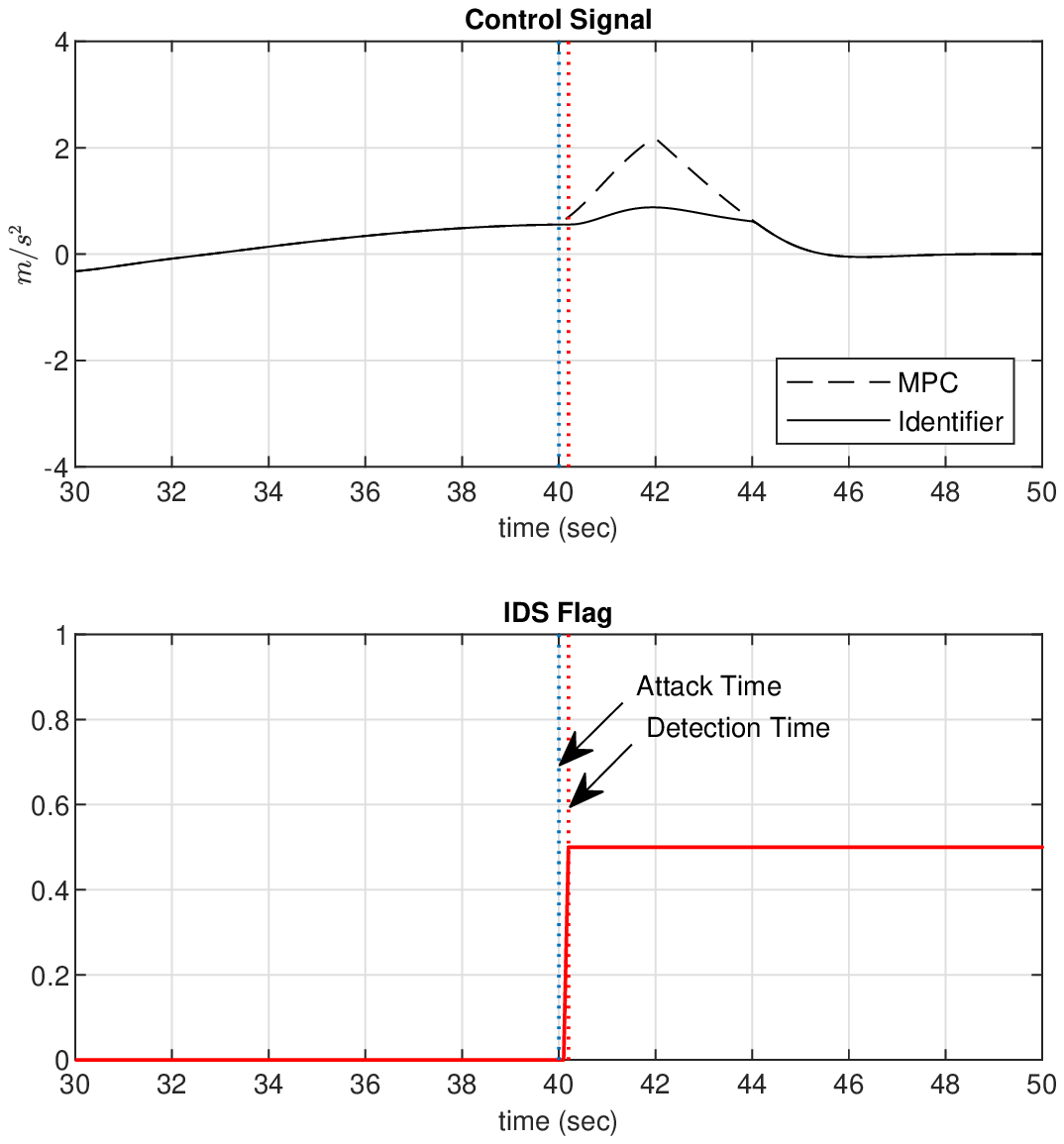}}
    %   \vspace{-2ex}
\caption{The intelligent intrusion detection system output for the two covert attack scenarios.  The attack time is   $t=40$ and the detection time is not much later than the attack time.  (a) The first attack scenario. (b) The second attack scenario.}
\label{fig.result_IDS}
%\end{left}
\end{figure*}

%\begin{figure}
%\includegraphics[width=\textwidth, angle=0, scale=.47]{Drawing1_1.eps}
%\caption{Representation of a 2D non-separable constellation with linear %support vector machine defining the boundary.}
%\label{fig.main}
%\end{figure*}
\begin{figure*}[!htb] %[b] %[p] puts at the end
%\begin{left}
%   \centering
   \subfloat[] {\includegraphics[width=1\columnwidth, height=1.2\columnwidth]{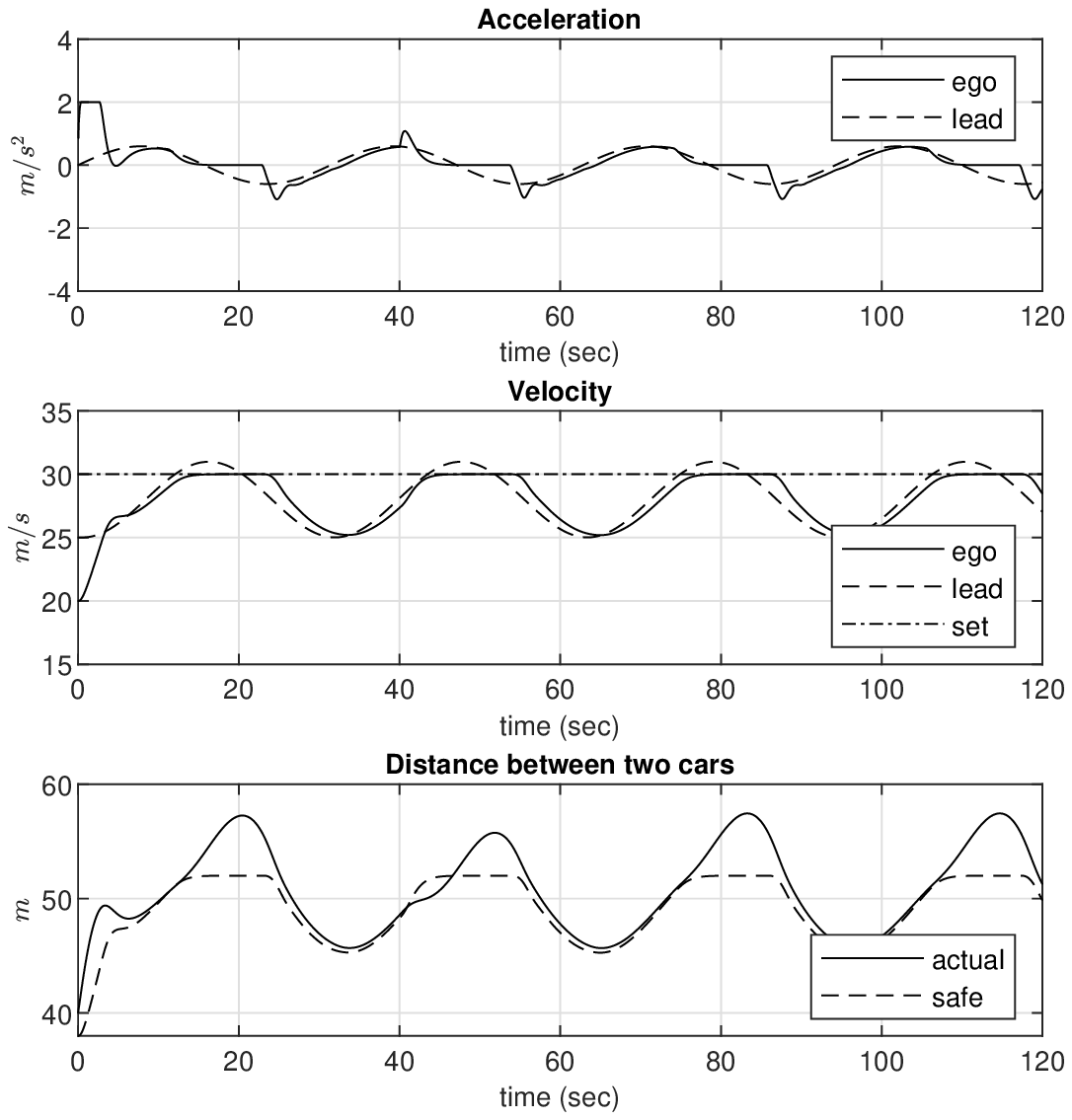}}
   \subfloat[]{\includegraphics[width=1\columnwidth, height=1.2\columnwidth]{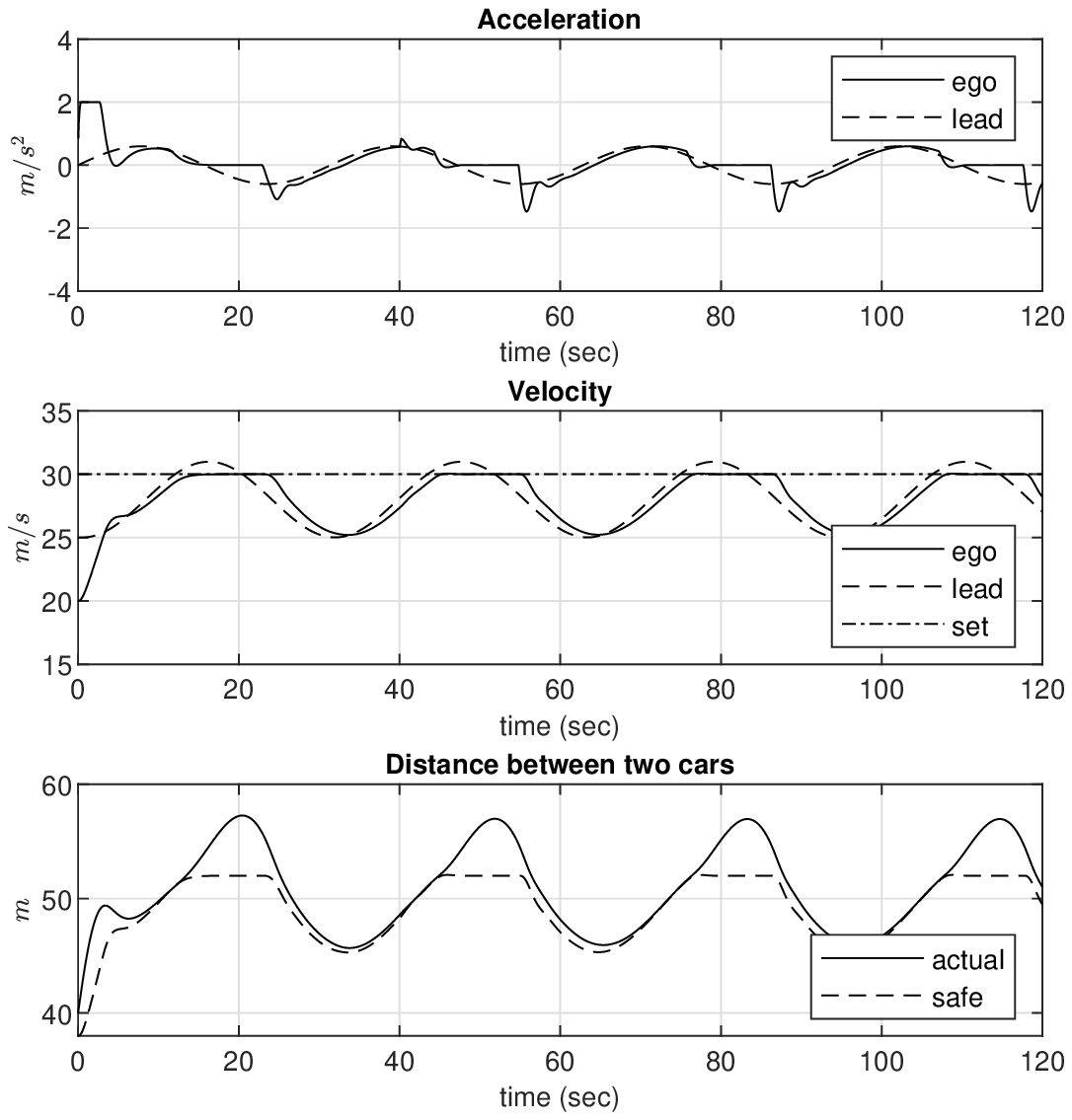}}
%   \vspace{-2ex}
\caption{The first attack scenario. (a) There is not any compensation strategy. (b) The P Controller is applied to compensate the malicious attack occurred into the ACC system.}
\label{fig.result_scen1}
%\end{left}
\end{figure*}
\begin{figure*}[!htb] %[b] %[p] puts at the end
%\begin{left}
%   \centering
   \subfloat []{\includegraphics[width=1\columnwidth, height=1.2\columnwidth]{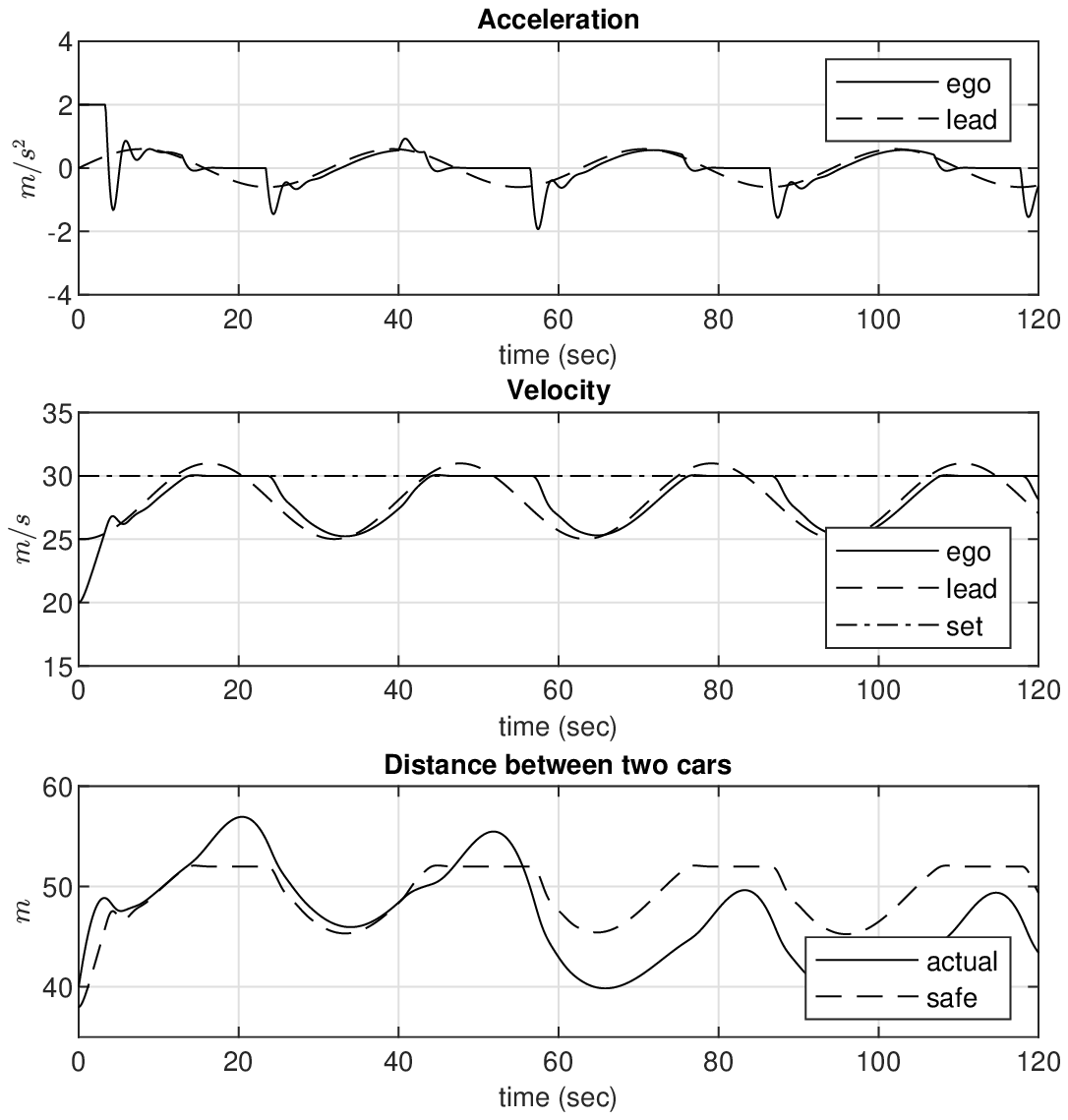}}
   \subfloat []{\includegraphics[width=1\columnwidth, height=1.2\columnwidth]{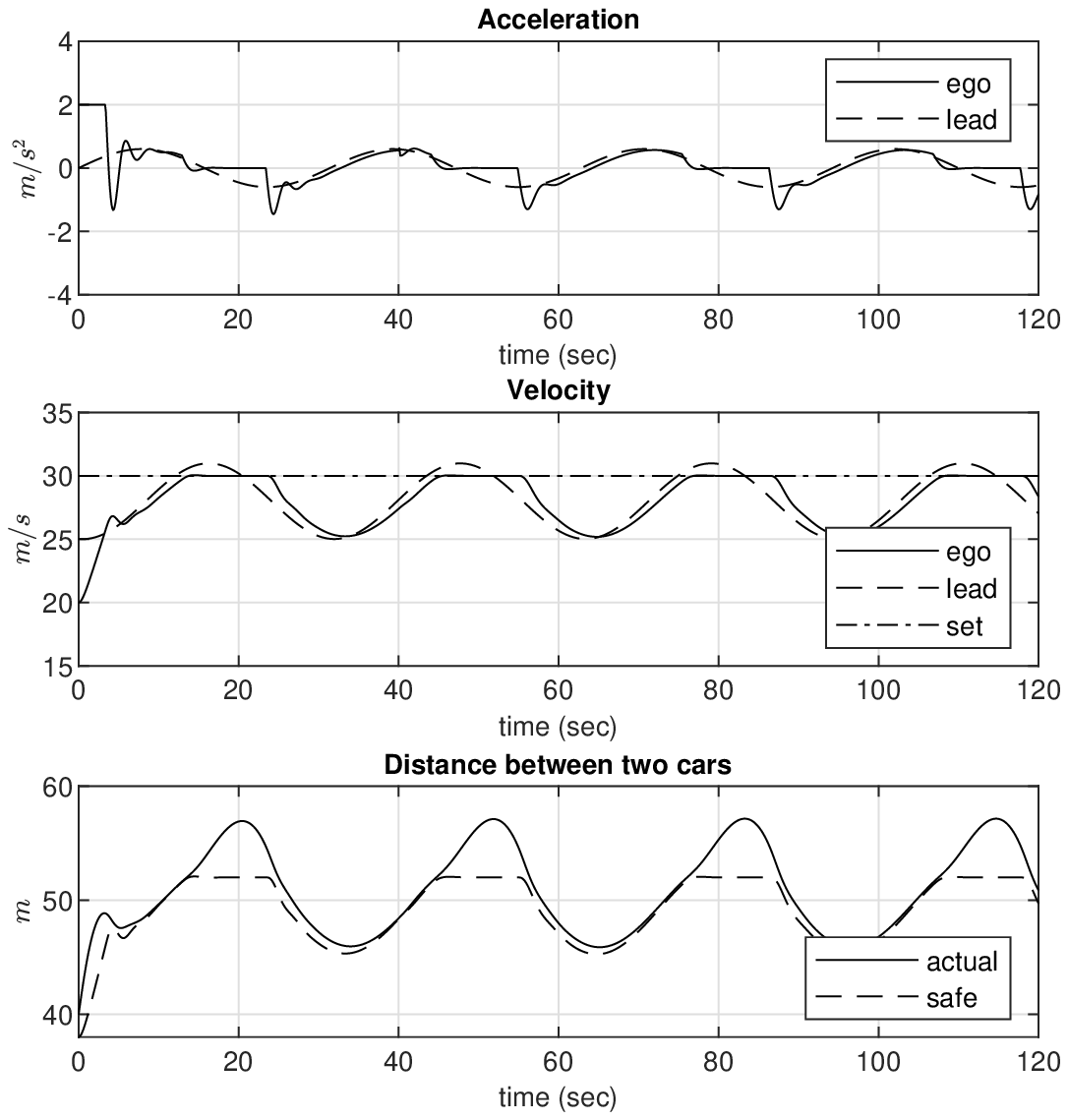}}
%   \vspace{-2ex}
\caption{The second attack scenario. (a) There is not any compensation strategy. (b) The P Controller is applied to compensate the malicious attack occurred into the ACC system.}
\label{fig.result_scen2}
%\end{left}
\end{figure*}

\section{Evaluation Results and Discussions\label{section:results}}

The proposed scheme for the ACC system   described in Section \ref{section:ACC}   is illustrated in Fig. \ref{fig:scheme}. As mentioned before, the ACC system is prone to malicious attacks. We introduced two types of covert attack scenarios  in subsection \ref{section:attack scenarios}.
In this section, the proposed strategies for detection and compensation  are applied to the smart vehicle system. The parameter values  used in the simulations are as follows:
The initial positions for the lead vehicle and the ego vehicle are $x_{0,lead} = 50~(m)$ and  $x_{0,ego} = 10~(m)$. 
The initial velocities for the lead vehicle and the ego vehicle are $v_{0,lead} = 25~(m/s)$ and  $v_{0,ego} = 20~(m/s)$.
Additionally, we assume that  
$Tgap=1.4~(s)$, $D_{default}=10~(m)$, $v_{set}=30~(m/s)$, $a_{ego,~ min}=-3~  (m/s^{2})$, and $a_{ego,~max}=2~(m/s^{2})$  .

\subsection{Intelligent Intrusion Detection System}
The results of applying the proposed intrusion detection system is shown in Fig.~\ref{fig.result_IDS}. The attack has been launched at $t=40~s$ in both scenarios. As observed, the  IDS alert flag is raised almost immediately. Therefore, the compensating system, which is a duplicate but simple replacement for the ACC unit, 
  is enabled after the detection of  covert attacks.

\subsection{Compensation of Covert Attacks - Attack Scenario \#1}
As mentioned before the first covert attack scenario is related to the speed control goal.  The attacker tries to  accelerate (the ego vehicle) when the lead  decreases its speed and the distance between the two cars gets lower, right at $t= 40s$ as shown in Fig.~\ref{fig.result_IDS} (a) and Fig. \ref{fig.result_scen1} (a). As a result, the gap between the two cars gets smaller than the safe  distance temporarily. This can be done repeatedly, however, only one attack instance has been shown in the figures.   The indicators  of the attack are visually hidden from the driver. The increase in acceleration and the reduction in car-to-car distance are not big enough to draw the attention of the ego diver, thus it remains stealth even after $t=40s$. However,  the probability of  accidents has been increased, at least in the  short  time window(s). With the IDS employed, the ACC system is taken over by an embedded P controller upon attack detection. The attack is detected shortly, after around $200ms.$ One can verify from Fig. \ref{fig.result_scen1} (b) that the  vehicle has managed to operate normally after $t=40.2s$ since IDS has taken the infected controller out of the loop and has put the embedded compensator in charge.    
\subsection{Compensation of Covert Attacks - Attack Scenario \#2}
As mentioned before, in the second covert attack scenario, the attacker tries to disrupt the reference of the ACC system, especially in mode 2, such that the gap between the two vehicles  remains below the safe distance. The IDS  reaction to this attack is  shown in Fig.~\ref{fig.result_IDS}(b). Since this is a covert attack, the driver of the ego vehicle does not necessarily sense any fluctuations in acceleration, as observed in Fig.~\ref{fig.result_scen2} (a). However, the relative distance gradually grows smaller and after around 60s, the effective gap between the two vehicles is reduced by almost 5m.
  This might not be noticeable (in 50m) by the driver but considerably increases the chance of accidents, especially when the lead car has good brakes and the ego car does not. 
Similar to the previous scenario, if  the embedded P compensator is enabled upon IDS signal, it manages to overcome the side effects of the attack. As the plots in Fig.~\ref{fig.result_scen2}(b) show, the attack is again detected in less than 200ms and the control goal is fairly satisfied after a small transient disturbance.

As the simulation results show, the proposed intrusion detection and compensation strategy is quick and effective. It can ensure  the relative distance between the ego and lead vehicles does not get smaller than the safe distance. 

\section{Related Work}\label{section:related work}
%According to the contributions of this paper, we discuss the related work in the following three subsections.

  We discuss the related work in some issues of intrusion detection and compensation in CPS and intelligent transportation systems, especially Vehicular Ad Hoc Networks (VANETs).

The issue of CPS security has received considerable attention ever since  Stuxnet \cite{falliere2011w32} struck. Intrusion detection in CPS and IoT systems has been realized by applying many classic and intelligent methods. In \cite{kang2016intrusion}, artificial neural networks were applied to detect intrusions which may happen in vehicular networks. In \cite{jokar2011specification}, an intrusion detection system (IDS) has been designed for Home Area Networks (HAN). The IDS is based on ZigBee technology, because ZigBee is dominant in HANs. Abnormal patterns of the network is detected using deviations from normal behavior. 

Deception attacks are considered instances of the man-in-the middle (MITM) attack, e.g. false data injection. False data injection attacks in electric power grids and wireless sensor networks are studied in \cite{liu2011false,deng2018false,mo2010false}. In these papers, the attackers are deemed to be aware of the system information  as well as the controller. Therefore, they proposed covert attacks on physical systems. In \cite{long2005denial} and \cite{teixeira2012revealing}, Packet delay and Denial of Service (DoS) attacks are modelled as stochastic processes in which the attacker launches zero-dynamic attacks on the system. Also, the stealth attacks for linear time invariant systems has been studied. Other viewpoints and algorithms for time-based IDS have been proposed in \cite{zimmer2010time}. Bound checking of execution micro-timings has been proposed to be adopted by application(s) in order to detect intrusions during a self test procedure. 

In \cite{de2017covert}, a covert attack has been studied for the purpose of service degradation. The goal was to study how covert attacks change the performance of networked control systems. In this paper, the attacker identifies the model of the plant and the controller. Then, the attacker determines what actions can damage the system, either in a short period or in the long run. Short-time attacks are designed to create overshoots on the system output and  long-term service degradation attacks aim creating a noticeable steady state errors. In \cite{farivardetection}, an intrusion detection and compensation system is proposed for CPS to fight covert attacks.  Errors of the output and its estimation are used for artificial neural network IDS. An intelligent controller is also designed to compensate the effect of attacks and it can replace the classic controller upon the attack detection. 
In \cite{Sayad2019CPS}, a hybrid strategy is applied to provide a tolerant control for compensation of cyber attacks launched on the inputs and outputs of a CPS of rotary gantry type.  The malicious attacks studied in this paper is Denial of Service (DoS) attacks which cause packet loss in the control input and  sensor output. In this paper, several classic and intelligent controllers have been studied in terms of robustness and effectiveness against cyber attacks.
The authors in \cite{sayad_firewall} took a different approach regarding the security of CPS. Instead of proposing an IDS, they showed how one can build an Intrusion Prevention System (IPS) for real-time cyber physical systems in general. Their approach results in a zero false-positive firewall whose rules are written automatically  by an algorithm from dataset. 

Intelligent transport systems are one of the major elements of  smart cities~\cite{xiong2012intelligent}. Vehicular Ad hoc Networks (VANETs) have become a key component of the intelligent transport system. VANETs provide safety information for both drivers and passengers.  There are many issues and limitations in VANETs  in practice  \cite{mokhtar2015survey,raya2006securing,darisini2013survey}. They include  network volatility, delay-sensitive applications, network scale, heterogeneity, wireless link usage, multi-hop connection, anonymity, etc. Some types of attacks affect the operation performance of VANETs \cite{mokhtar2015survey,djenouri2005security}. Attacks are categorized into  insider (launched by internal authorized malicious vehicles) and external (launched from outsiders). From another perspective, attacks are classified as passive and active. Reference  \cite{hasrouny2017vanet} also classifies attacks into four main groups which pose (1) a risk to wireless interface, (2) a threat to hardware and software, (3)  a hazard to sensors in-put in vehicle and (4) a danger behind wireless access (in the infrastructure). Malicious attacks in VANETs can be message spoofing, message replay attack, integrity attack, impersonation attack, Denial of service (DoS) attack, De-anonymization attack \cite{haghighi2019highly}, movement tracking, etc.

In \cite{jolfaei2019privacy}, data security and privacy issues
in intelligent transportation systems have been studied. In this paper, it has been assumed that data streams
are coming out of  vehicles and go to road side units. A group formation in the vehicular layer is proposed in which the (group) leader can communicate with the members as well as the road side unit. A lightweight permutation mechanism is applied to preserve the
confidentiality and privacy of sensory data.

In \cite{kumar2018detection}, an algorithm is presented  that can detect jamming attacks of DoS type in VANETs. In the proposed detection algorithm,  multiple malicious nodes and irrelevant nodes can be detected and isolated from the routing network. In \cite{liang2019novel}, an IDS  is proposed which can be used in the wireless and dynamic networks, like VANETs. A novel algorithm, which contains a feature extraction method as well as a  hierarchical classifier, has been  developed  for intrusion detection in VANETs.  The "differences of traffic flow" and  "position" are the two main features  that should be extracted. The classifier works based on relabeling and recalculating mechanisms.

\section{Conclusion\label{section:conclusion}}

In this paper, detection and compensation of covert intrusions into adaptive car cruise control  system (ACC) were studied. Two scenarios for covert attacks on ACC were introduced  such that the ACC system is not able to satisfy the speed and space control goals of the smart vehicle.
An artificial neural network identifier was proposed to learn the ACC system and predict its outputs. The IDS system works based on comparison of the actual ACC outputs with those of the identifier that received the same inputs. Anomalies are captured by statistical measures and they raise a flag  that switches the MPC system to an embedded PID controller. 
Simulation results confirmed that the proposed approach was effective in the sense that it both achieved detecting the covert attacks and mitigated their effects on the performance of  tested vehicle.  

\section*{Acknowledgement}
Authors are thankful to the Advanced Networking and Security research Laboratory (ANSLab.org) for the supports provided during this study.

\bibliographystyle{IEEEtran}
{\small \bibliography{references}}

\end{document}